\documentclass{aastex}
\usepackage{emulateapj5} 

\bibpunct{(}{)}{;}{a}{}{,}

\usepackage{txfonts}
\usepackage{rotating}
\usepackage{color}
\usepackage{multirow}
\usepackage{ulem}

\newcommand{\grb}{GRB\,111005A}
\newcommand{\hi}{\sc Hi}
\newcommand{\kms}{\mbox{km\,s$^{-1}$}}

\newcommand{\inst}[1]{\altaffilmark{#1}}
\newcommand{\tablebib}{\tablecomments}

\newcommand\revone[1]{#1}
\newcommand\revtwo[1]{#1}
\newcommand\revthree[1]{#1}
\newcommand\revfour[1]{#1}

\shorttitle{The ISM in the environment of GRB\,111005A} 
\shortauthors{Le\'sniewska et al.} 

\begin{document}

   \title{
   The interstellar medium in the environment of the supernova-less long-duration GRB\,111005A
  }

\author{
Aleksandra Le\'sniewska\inst{\ref{inst:poz}},
M.~J.~Micha{\l}owski\inst{\ref{inst:poz}, \ref{inst:edi}, \ref{inst:cal}, \ref{inst:fulb}},
P.~Kamphuis\inst{\ref{inst:rub}},
K.~Dziadura\inst{\ref{inst:poz}},
M.~Baes\inst{\ref{inst:ugen}},
J.~M.~Castro~Cer\'{o}n\inst{\ref{inst:esac}},
G.~Gentile\inst{\ref{inst:vrije}},
J.~Hjorth\inst{\ref{inst:dark}},
L.~K.~Hunt\inst{\ref{inst:hunt}},
C.~K.~Jespersen\inst{\ref{inst:dawn}}
M.~P.~Koprowski\inst{\ref{inst:umk}},
E.~Le Floc'h\inst{\ref{inst:sacley}},
H.~Miraghaei\inst{\ref{inst:riaam}},
A.~Nicuesa Guelbenzu\inst{\ref{inst:taut}}, 
D.~Oszkiewicz\inst{\ref{inst:poz}},
E.~Palazzi\inst{\ref{inst:pal}},
M.~Poli\'nska\inst{\ref{inst:poz}},
J.~Rasmussen\inst{\ref{inst:dtu}},
P.~Schady\inst{\ref{inst:bath}},
D.~Watson\inst{\ref{inst:dawn}}
    }

\altaffiltext{1}
{Astronomical Observatory Institute, Faculty of Physics, Adam Mickiewicz University, ul. S{\l}oneczna 36, Pozna{\'n}, Poland
\label{inst:poz}}

\altaffiltext{2}
{Scottish Universities Physics Alliance (SUPA), Institute for Astronomy, University of Edinburgh, Royal Observatory, Blakford Hill, EH9 3HJ, Edinburgh, UK
\label{inst:edi}}

\altaffiltext{3}
{TAPIR, Mailcode 350-17, California Institute of Technology, Pasadena, CA 91125, USA
\label{inst:cal}}

\altaffiltext{4}
{Fulbright Senior Award Fellow
\label{inst:fulb}}

\altaffiltext{5}
{Ruhr-Universit\"at Bochum, Faculty of Physics and Astronomy, Astronomical Institute, 44780 Bochum, Germany
\label{inst:rub}}

\altaffiltext{6}
{Sterrenkundig Observatorium, Universiteit Gent, Krijgslaan 281-S9, 9000, Gent, Belgium
\label{inst:ugen}}


\altaffiltext{7}
{ISDEFE for the {\em SMOS} FOS (ESA-ESAC), E-28.692 Villanueva de la Ca\~nada (Madrid), Spain
\label{inst:esac}}

\altaffiltext{8}
{Department of Physics and Astrophysics, Vrije Universiteit Brussel, Pleinlaan 2, 1050, Brussels, Belgium
\label{inst:vrije}}

\altaffiltext{9}
{DARK, Niels Bohr Institute, University of Copenhagen, Jagtvej 128, DK-2200 Copenhagen N, Denmark
\label{inst:dark}}

\altaffiltext{10}
{INAF-Osservatorio Astrofisico di Arcetri, Largo E. Fermi 5, I-50125 Firenze, Italy
\label{inst:hunt}}

\altaffiltext{11}
{The Cosmic Dawn Center, Niels Bohr Institute, University of Copenhagen, Lyngbyvej 2, DK-2100 Copenhagen \O, Denmark
\label{inst:dawn}}

\altaffiltext{12}
{Institute of Astronomy, Faculty of Physics, Astronomy and Informatics, Nicolaus Copernicus University, Grudzi\c{a}dzka 5, 87-100 Toru\'{n}, Poland
\label{inst:umk}}

\altaffiltext{13}
{Laboratoire AIM-Paris-Saclay, CEA/DSM/Irfu - CNRS - Universit\'e Paris Diderot, CE-Saclay, pt courrier 131, F-91191 Gif-sur-Yvette, France
\label{inst:sacley}}

\altaffiltext{14}
{Research Institute for Astronomy and Astrophysics of Maragha (RIAAM), University of Maragheh, Maragheh, Iran
\label{inst:riaam}}

\altaffiltext{15}
{Th\"uringer Landessternwarte Tautenburg, Sternwarte 5, 07778 Tautenburg, Germany
\label{inst:taut}}

\altaffiltext{16}
{INAF-OAS Bologna, Via Gobetti 93/3, I-40129 Bologna, Italy
\label{inst:pal}}

\altaffiltext{17}
{Technical University of Denmark, Department of Physics, Fysikvej, building 309, DK-2800 Kgs. Lyngby, Denmark
\label{inst:dtu}}

\altaffiltext{18}
{Department of Physics, University of Bath, Bath, BA2 7AY, United Kingdom
\label{inst:bath}}

\date{Received X; accepted X}

\begin{abstract}

Long ($>2$\,s) gamma ray bursts (GRBs) 
are associated with explosions of massive stars, although in three instances, supernovae (SNe) have not been detected, despite deep observations. 
\revone{With new {\hi} line and archival optical integral field spectroscopy data}, we characterize the interstellar medium (ISM) of the host galaxy of one of these events, {\grb}, in order to shed light on the unclear nature of these peculiar objects.

We found that the atomic gas, radio continuum, and rotational patterns are in general very smooth throughout the galaxy, which does not indicate a recent gas inflow or outflow. There is also no gas concentration around the GRB position. The ISM in this galaxy differs from that in hosts of other GRBs and SNe, which may suggest that the progenitor of {\grb} was not an explosion of a very massive star (e.g.~a compact object merger).

However, there are subtle irregularities of the {\grb} host \revone{(most at a $2\sigma$ level)}, which may point to a weak gas inflow or interaction. 
\revone{Since in the SE part of the host there is 15\% more atomic gas and twice less molecular gas than in NW part, the molecular gas fraction is low.}
In the SE part there is also a region with very high H$\alpha$ equivalent width. There is more continuum 1.4\,GHz emission to the SE and an S-shaped warp in the UV. Finally, there is also a low-metallicity region 3.5{\arcsec} (1\,kpc) from the GRB position. Two  galaxies  within 300\,kpc or a past merger can be responsible for these irregularities.

\end{abstract}

\keywords{Supernovae (1668); Interstellar atomic gas (833); HI line emission (690); Spiral galaxies (1560); Galaxy kinematics (602); Star formation (1569); Gamma-ray bursts (629)}

\section{Introduction}
It is well established that there are two types of gamma-ray bursts (GRBs) and they are divided based on the duration of their prompt emission \citep{Kouveliotou93}. Those lasting less than 2 seconds are called short GRBs, with the cause of the explosion being the collision of compact objects \citep[two neutron stars or a neutron star and a black hole;][]{Abbott2017}. The second group, longer than 2 seconds (long GRBs) are the result of the  core collapse of very massive stars \citep{Hjorth2003,Stanek2003} and
take place in galaxies with ongoing star-formation \citep{Christensen2004,Castroceron2006,Castroceron2010,Michalowski2008,Savaglio2009,Perley2013,Perley2015,Hunt2014}.

Almost all long GRBs for which deep spectroscopic observations were carried out are accompanied by the explosion of supernovae type Ic (SN Ic; with no hydrogen, helium, or silicon lines in the spectrum; \citealt{Hjorth2012}).
There are three exceptions for which the existence of a SN was ruled out down to deep limits:  GRB\,060505, 060614 \citep{Fynbo2006,Dellavalle2006,Galyam2006}, and 111005A \citep{Michalowski2018grb}.

The most important question regarding these three objects is what their nature and their progenitors were.  Some clues on the nature of unusual explosions can be obtained from the properties of gas in their environment \citep[as was done for the enigmatic transient AT\,2018cow;][]{Michalowski2019,Roychowdhury2019,Morokumamatsui2019}. In this paper we focus on {\grb}. The current gas data for its host have too poor resolution to attempt this.
\cite{Michalowski2018co} obtained a detection of the CO(2-1) line in three pointings and noticed that ESO 580-49, the host galaxy of {\grb}, is not symmetrically filled with molecular gas. The central and NW regions are  molecule-rich for their star formation rates (SFR), but the SE region turned out to be molecule-deficient. 
The total molecular gas mass turned out to be similar as that of other galaxies with similar redshift, star formation rate (SFR), and stellar mass \citep{Michalowski2018co,Hatsukade2020}.
\cite{Michalowski2015} examined the {\hi} line in five GRB host galaxies at \textit{z} $<$ 0.12, including the {\grb} host using  archival {\hi}
line from the Nan\c{c}ay radio telescope \citep{Theureau1998,Springob2005}. Due to poor spatial resolution, only the total atomic gas mass $\mbox{M}_{\rm HI,tot,single-dish}$ was measured, and it turned out to be typical for the SFR and stellar mass of this galaxy. 

The objective of this paper is to establish the properties of the interstellar medium (ISM) in the host of {\grb}, in order to help shed light on the nature of the progenitor.
We use new {\hi}  and archival integral spectroscopy data to investigate both the atomic and the ionised gas.

We use a cosmological model with $H_0$ = 70 km s$^{-1}$ Mpc$^{-1}$, $\Omega_{\Lambda}$~=~0.7, and $\Omega_m$~=~0.3. At the redshift of {\grb} of $0.01326$ this corresponds to a scale of 0.27 kpc per 1\arcsec .

\section{GRB\,{\scriptsize 111005}A and its host}

On October 5$^{th}$ 2011 {\grb} was detected by the Burst Alert Telescope (BAT, \citealt{Barthelmy2005}), onboard the \textit{Swift} satellite. With a burst duration of 26 $\pm$ 7 sec \citep{Barthelmy2011}, it was classified as a long GRB. 
\revone{Machine-learning classification that successfully distinguishes between long and short GRBs, shows that {\grb} is in the long GRB part of the diagram, in common with other supernova-less long GRBs as well \citep{Jespersen2020}. In this analysis, {\grb} lies far from the short-duration group in a densely populated area, making it unlikely that it could belong to the short-duration group.
This is consistent with its duration, putting it in the long category.}

\cite{Michalowski2018grb} carried out an analysis of the GRB afterglow and explosion environment based on new and archival radio, optical and mid-infrared data. With Very Long Baseline Array (VLBA) data the position of the radio afterglow was determined to be $\alpha$~=~14:53:07.8078276, $\delta$ = $-$19:44:11.995387 (J2000), with a 1$\sigma$ error of 0.2 mas.
The GRB was found to be associated with the galaxy ESO 580-49 at a redshift of \textit{z} = 0.01326, exploding $\sim1''$ from its centre, \revone{as defined on the 3.6\,{\micron} image \citep{Michalowski2018grb}}.
The radio  afterglow lightcurve 
turned out to be atypical.
The afterglow exhibited a plateau phase lasting a month with a very rapid subsequent decay. These properties have never been observed before in a GRB. The host galaxy has been classified as Sbc in HyperLeda\footnote{\url{leda.univ-lyon1.fr}} with an edge-on inclination of 90$^{\circ}$ \citep{hyperleda}. Based on full spectral energy distribution (SED) modelling including far-infrared data, \cite{Michalowski2018grb} determined a star formation rate (SFR) of 0.42~$_{-0.05}^{+0.06}$ $M_{\odot} yr^{-1}$ and stellar mass of $\log (M_{*}/M_{\odot})= 9.68$ $_{-0.09}^{+0.13}$ which is  within the range of both long and  short GRBs hosts  \citep{Savaglio2009,Castroceron2010,Perley2016,Fong2013,Berger2014,Klose2019}.
\cite{Michalowski2018grb} and \cite{Tanga2018} rejected the characterisation of the host galaxy as an active galactic nucleus (AGN) based on the Baldwin-Phillips-Terlevich (BPT) diagram \citep{bpt}.

Another rare feature of {\grb} was that  no supernova was detected in the optical, near- and mid-infrared, down to an absolute magnitude of $-12$\,mag \revone{at $3.6\,\micron$}, $\sim20$ times fainter than SNe associated with long GRBs. This cannot be explained by dust extinction, because in the mid-infrared its influence would be minor.
This is similar to GRBs 060505 and 060614 \citep{Fynbo2006,Dellavalle2006,Galyam2006}.

Moreover, \cite{Michalowski2018grb} showed that the explosion occurred in an environment with about solar metallicity. A very similar conclusion was reached by \cite{Tanga2018} based on integral field spectroscopy. They found that the host galaxy is metal-rich (near solar metallicity) and that there is little star formation at the GRB position. There are \revtwo{about twenty known}  
GRBs that have exploded in environments with high (solar \revone{or super-solar}) metallicity, \revone{(measured from absorption spectroscopy) 
} \citep{Prochaska2009,Kruhler2012,Kruhler2015,Savaglio2012,Elliott2013,Schulze2014,Hashimoto2015,Schady2015,Stanway2015b}.
\revtwo{Metallicity is dependent on the galaxy stellar mass, so massive GRB hosts without metalliticty measurements may also have high metallicity. \citet{Perley2016} presented an analysis of 119 galaxies up to redshift 6, of which about 10$\%$ are massive galaxies (stellar masses larger than $10^{10.5}$), which imply solar metallicity. However, all of them are at $z>1$, whereas the {\grb} host is at a low redshift.}
\revone{GRB\,130925A was also similar to {\grb} with respect to its proximity to the host centre \citep[$0.12\arcsec$ or 600\,pc in projection;][]{Schady2015}.}
The lack of a SN, an atypical lightcurve, and the high (around solar) metallicity 
\revone{were used to claim that the explosion mechanism of {\grb} was different from that of the majority of GRBs \citep{Michalowski2018grb,Tanga2018}. 
}

\section{Data}

\subsection{GMRT observations}

In May and June 2016 the field of {\grb} was observed for 2 $\times $ 3 hrs with the Giant Metrewave Radio Telescope (GMRT)\footnote{Project no.~30$\_$035, PI: M.~Micha{\l}owski}. For calibration of the flux and the bandpass 3C286 was observed for 15 min at the start and the end of the runs. For the phase calibration 1522-275  was observed every 30 min on one day and 1448-163 the other day with the same temporal spacing. The correlator was setup with 16 Mhz bandwidth and 512 channels centred around 1400 \revone {MHz}.

The data were reduced with a range of data reduction packages. We downloaded the fits files with the raw data from the GMRT archive. These fits files were then loaded into {\sc CASA} \citep{McMullin2007} with the {\sc importgmrt} task without applying the online flags. Further data reduction was done with the \revone{{\sc CARACal}\footnote{\url{https://github.com/caracal-pipeline/caracal}}} pipeline which is being developed for H{\sc i} data reduction of MeerKAT data. The pipeline is setup in a modular fashion using the platform-independent radio interferometry scripting framework {\sc stimela}\footnote{\url{https://github.com/SpheMakh/Stimela/wiki}}. In practice this means that the calibrator data  are initially flagged with {\sc AOflagger} \citep{Offringa2010} and calibrated and transferred to the target with {\sc CASA}. After this the target is split out of the measurement set, further flagged with {\sc AOflagger}, imaged with {\sc WSclean} \citep{Offringa2014} in Stokes I, using {\sc WSclean}'s auto and fits masking feature after which this clean model is  
\revone {used in} {\sc Cubical}\footnote{\url{https://github.com/ratt-ru/CubiCal}} \citep{Kenyon2018} for the self-calibration. This step is repeated until a phase-only self-calibration no  longer improves the image and then subsequently an 
\revone {amplitude and phase} self-calibration is performed where the  solution interval for the amplitude and phase can differ.

After calibration the data for the two separate days were mapped onto the same channel grid with the {\sc CASA} task {\sc mstransform} and the continuum was subtracted with {\sc uvcontsub}. At this stage the data were also Doppler corrected and projected onto a barycentric velocity frame. As the data for the two separate days had opposite sign frequency increment, the pipeline's more advanced {\sc WSclean} tasks could not be used to invert the visibilities into an H{\sc i} data cube. For this reason we did this final step manually in {\sc CASA}. {\sc TCLEAN} was used to transform the visibilities of the 60 channels covering the H{\sc i} emission into a data cube. The visibilities were weighted according to a Briggs weighting scheme with the ``robust'' parameter set to 0 and a uvtaper of 7, 17, and 40 k$\lambda$ resulting in data cubes of different resolution.
The dirty cubes were cleaned with {\sc TCLEAN}'s multiscale clean algorithm, at scales of 1, 2  and 5 beams. The cleaning was performed in an iterative process in which we first cleaned the cube to a threshold of 10$\sigma$ \revone{in the first iteration. From this first ``cleaned" cube a mask was constructed with {\sc SoFiA} \citep{Serra2015} and then the emission in this mask was cleaned down to  0.5$\sigma$ \revtwo{and a new mask was created}. This last step  was repeated until successive iterations showed no changes in the mask and all the visible emission was captured in the mask.}

The final cubes had resolutions of FWHM = 16.8\arcsec$\times$12.9\arcsec, 7.4\arcsec$\times$5.7\arcsec, and 4.0\arcsec$\times$2.8\arcsec and a channel width of 32.5 kHz (7\,{\kms} at the {\hi} frequency).
The rms is $1.0$, $0.7$, and $0.6\,\mbox{mJy\,beam}^{-1}$ per channel.
The frequency axis was converted into a velocity axis using the relativistic definition which results in a channel width of 6.96 km s$^{-1}$ with an error of $\sim$ 0.005 km\,s$^{-1}$ on the outermost channels of the cube. 

We also imaged the line-free channels to construct continuum images at the frequency of 1.4\,GHz. We applied a uvtaper of 40, 17 and 5k$\lambda$ resulting in a resolution of 4.7\arcsec$\times$3.2\arcsec,  8.4\arcsec$\times$6.5\arcsec and \revone{29.7\arcsec$\times$22.5\arcsec} and an rms of  \revone{$40$, $66$ and $178\,\mu\mbox{Jy\,beam}^{-1}$}, respectively. \revone{For the flux scale we note that the usual 10\% calibration error applies for GMRT data.}

\subsection{MUSE data}

We also used optical data obtained by \cite{Tanga2018}.
The galaxy in which {\grb} occurred was observed on the 23 August 2014 by the ESO Very Large Telescope equipped with the  panoramic integral-field spectrograph working in the visible range, the Multi-Unit Spectroscopic Explorer (MUSE; \citealt{Bacon2010}).

The MUSE optical data obtained by \cite{Tanga2018} has previously been used in order to study the dust reddening $E(B-V)$, velocity map based on the H$\alpha$ line, equivalent width of the H$\alpha$ line, star formation rate surface density, and metallicity based on the \citet[][O3N2]{Pettini2004} and \citet[][D16]{Dopita2016} calibrations.

\begin{figure*}
\includegraphics[width=\textwidth]{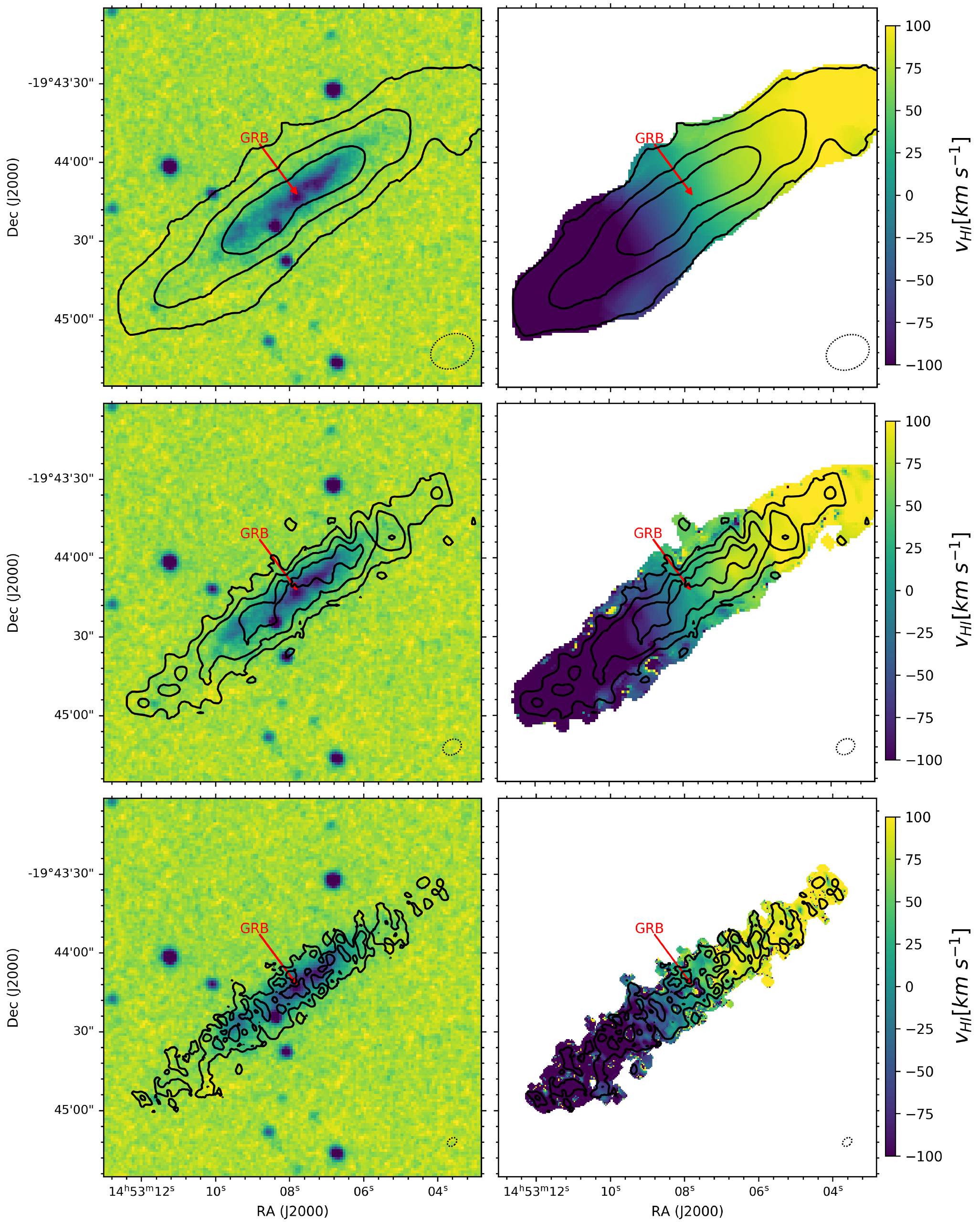}
\caption{\revone{(Left) {\hi} line emission of the {\grb} host (black contours) detected by GMRT with three different resolutions from top to bottom: 16.8\arcsec$\times$12.9\arcsec, 7.4\arcsec$\times$5.7\arcsec, and 4.0\arcsec$\times$2.8{\arcsec} (the beams are shown as grey dotted ellipses). 
The contours start at $0.7$, $0.04$, and $0.03\,\mbox{Jy\,beam}^{-1}\,\kms$ from top to bottom. The lowest contours correspond to column densities of $3.7$, $1.1$, and $3.0\times10^{21}\,\mbox{cm}^2$.
The background is the UV image from \citet{Michalowski2018grb}.
(Right) {\hi} 1$^{st}$ moment maps, with respect to the redshift \textit{z} = 0.01326. The position of {\grb} is marked by the red arrow. The image size is 2.5\arcmin$\times$2.5\arcmin corresponding to 40.5 kpc $\times$ 40.5 kpc.}}
\label{HI}
\end{figure*}

\begin{figure*}
\centering
\includegraphics[width=\textwidth]{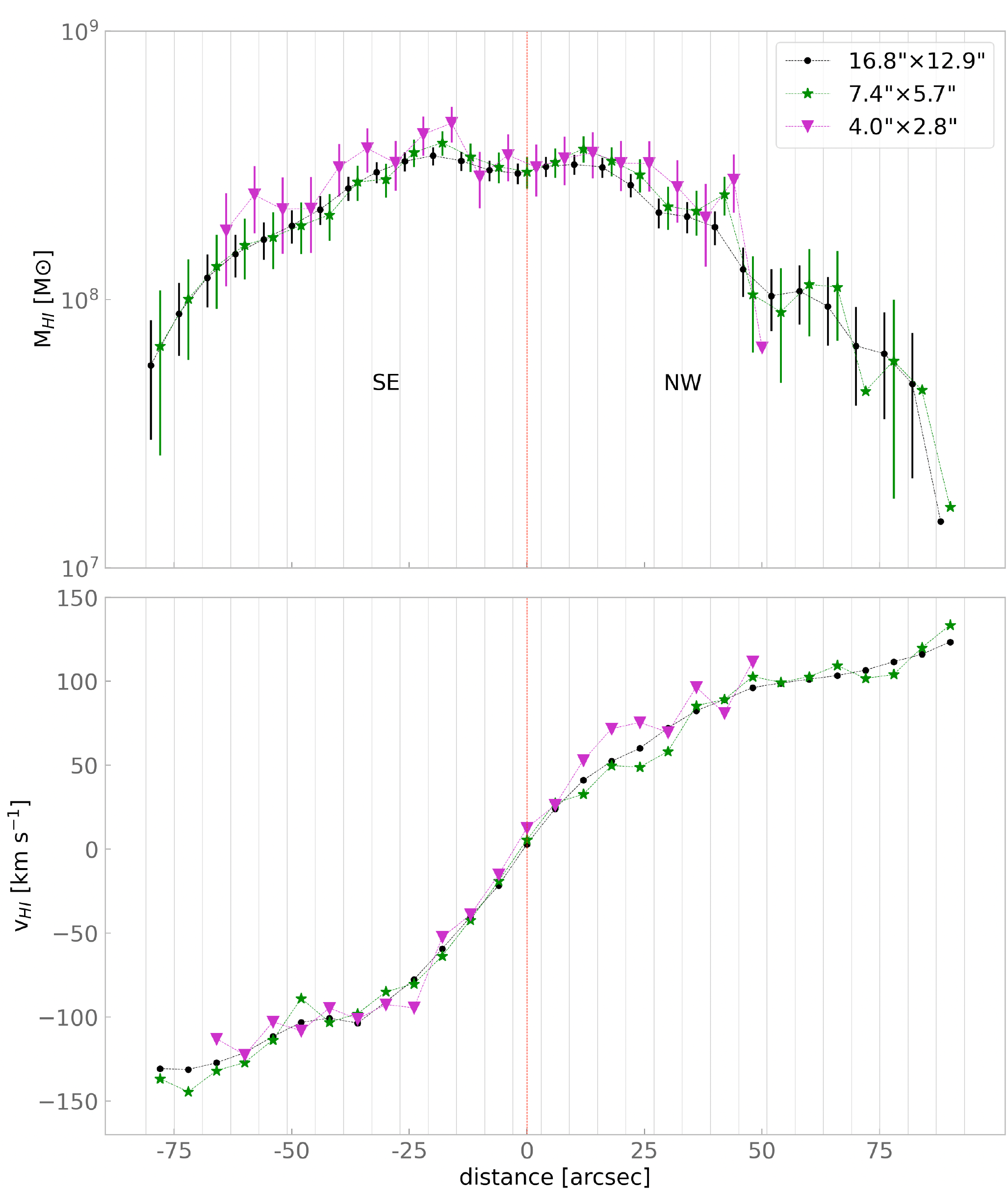}
\caption{\revone{Profiles of the atomic gas mass derived from the {\hi} maps shown on Fig. \ref{HI}, as a function of distance from the galaxy centre (top) and the rotation curve from the velocity field of {\hi} (bottom). For both panels the galaxy centre is marked as the red dashed line. The points on the top panel for which uncertainties are not shown represent upper limits.}}
\label{HIProfiles}
\end{figure*}

\begin{figure*}
\includegraphics[width=\textwidth]{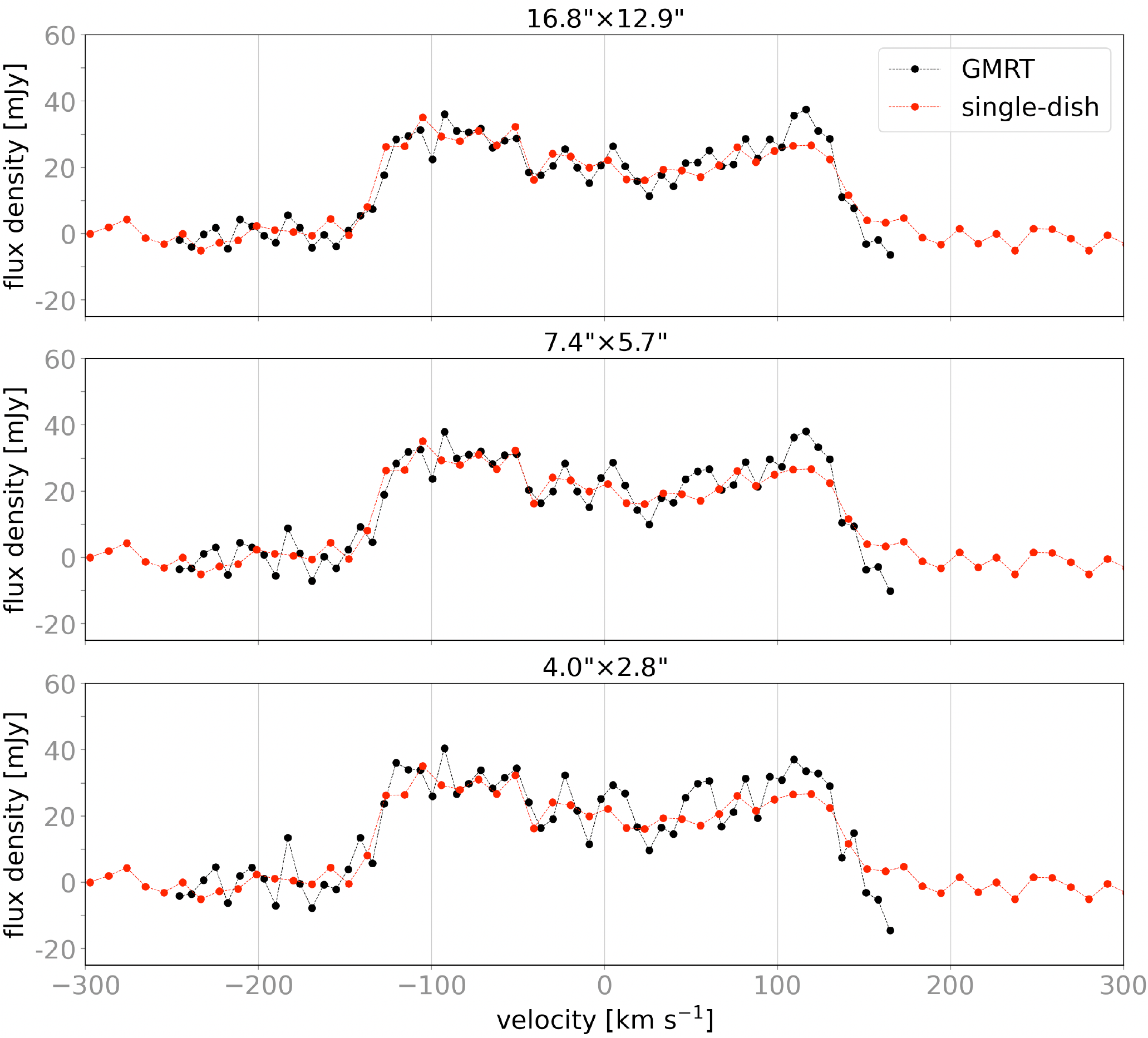}
\caption{\revone{GMRT interferometric spectrum (black dots) at three different resolutions and the single-dish spectrum (red dots; \citealt{Theureau1998}). The velocity axis is with respect to the redshift \textit{z} = 0.01326.}}
\label{spectrum}
\end{figure*}

\begin{table*}
\small
\caption{Properties of regions in the {\grb} host (Center, NW, and SE; see Fig.~\ref{apertures}) and of the entire galaxy.}
\centering
\begin{tabular}{ccccccccccccccc}
\hline\hline
\multirow{2}{*}{Reg.} & \multirow{2}{*}{RA/Dec} & &  \multicolumn{3}{c}{M$_{\rm HI}$} & \multicolumn{3}{c}{M$_{\rm HI}/\mbox{M}_{\rm HI,tot}$} & \multicolumn{3}{c}{molecular} & \multicolumn{2}{c}{$Z$} &  SFR \\
 &  &  & \multicolumn{3}{c}{(10$^9$ $M_\odot$)} & \multicolumn{3}{c}{(\%)} & \multicolumn{3}{c}{gas fraction (\%)} & \multicolumn{2}{c}{{$12 + \log(O/H)$}} & ($M_\odot$ yr$^{-1}$) \\
\cline{4-14}
 & &  & {14.85\arcsec} &{6.55\arcsec} &{3.4\arcsec} &{14.85\arcsec} &{6.55\arcsec} &{3.4\arcsec}&{14.85\arcsec}&{6.55\arcsec} &{3.4\arcsec} & D16 & O3N2 & \\
\hline
Center & 14:53:7.8 & & \revthree{0.90} & \revthree{1.08} & \revthree{1.17} & \revthree{16.0} & \revthree{18.8} & \revthree{19.4} & \revthree{38.2} & \revthree{34.3} & \revthree{32.5} & 8.39 & 8.55 & 0.12 \\ 
 & -19:44:10.97 & \revone{$\pm$} & \revthree{0.02} & \revthree{0.03} & \revthree{0.03} & \revthree{0.7} & \revthree{0.8} & \revthree{0.8} & \revthree{1.8} & \revthree{1.6} & \revthree{1.6}\\
 
NW & 14:53:6.2 & & \revthree{0.80} & \revthree{0.95} & \revthree{1.03} & \revthree{14.1} & \revthree{16.5} & \revthree{17.1} & \revthree{20.4} & \revthree{17.7} & \revthree{16.5} & 8.23 & 8.49 & 0.03\\
 & -19:43:56.03 & \revone{$\pm$ } & \revthree{0.02} & \revthree{0.02} & \revthree{0.03} & \revthree{0.6} & \revthree{0.7} & \revthree{0.8} & \revthree{1.6} & \revthree{1.8} & \revthree{2.0}\\
 
SE & 14:53:9.2 & & \revthree{0.96} & \revthree{1.16} & \revthree{1.25} & \revthree{17.0} & \revthree{20.3} & \revthree{20.8} & \revthree{10.0} & \revthree{8.4} & \revthree{7.9} & 8.38 & 8.45 & 0.04\\
 & -19:44:29.04 & \revone{$\pm$ } & \revthree{0.03} & \revthree{0.03} & \revthree{0.03} & \revthree{0.7} & \revthree{0.9} & \revthree{0.9} & \revthree{1.7} & \revthree{2.4} & \revthree{2.2}\\
 
Total & & & \revthree{5.66} & \revthree{5.73} & \revthree{6.01} & \revthree{94.4$^{\dagger}$} & \revthree{95.6$^{\dagger}$} & \revthree{100.3$^{\dagger}$} & \revthree{24.7} & \revthree{21.5} & \revthree{20.2} & 8.37 & 8.51 & 0.19\\
 & & \revone{$\pm$ } & \revthree{0.19} & \revthree{0.20} & \revthree{0.22} & \revthree{4.4} & \revthree{4.6} & \revthree{4.9} & \revthree{1.5} & \revthree{1.5} & \revthree{1.5}\\
\hline
\end{tabular}
\label{tab1}
\tablebib{Atomic gas masses (M$_{\rm HI}$), percentages of M$_{\rm HI}$ relative to the total mass of M$_{\rm HI}$ in the host galaxy \revone{(from the `Total' row)}, \revone{percentages of} molecular gas fractions ($M_{\rm H2}/M_{\rm gas}$) based on the H$_2$ masses obtained by \cite{Michalowski2018co}, metallicities determined from the \citet[][D16]{Dopita2016} and \citet[][O3N2]{Pettini2004} methods, and H$\alpha$ based star formation rates. The {\hi} properties are given in three columns corresponding to three different resolutions, indicated in arcsec. \newline
${\dagger}$ indicates the ratio between the total amount of atomic gas from the GMRT data and the total amount of atomic gas measured by \cite{Michalowski2015} based on single-dish data.}
\end{table*}

\begin{figure*}
\centering
\includegraphics[width=\textwidth]{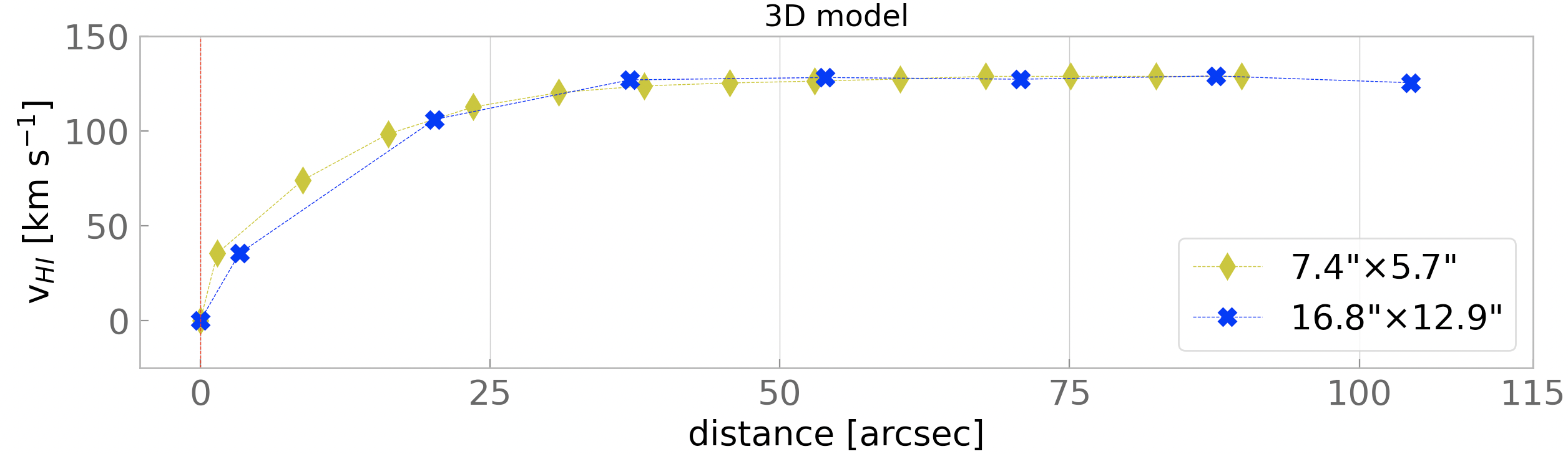}
\caption{\revone{Rotation curves from the tilted ring modeling at two different resolutions as a function of distance from the galaxy centre.}}
\label{3Dmodel}
\end{figure*}

\begin{figure*}
\includegraphics[width=\textwidth]{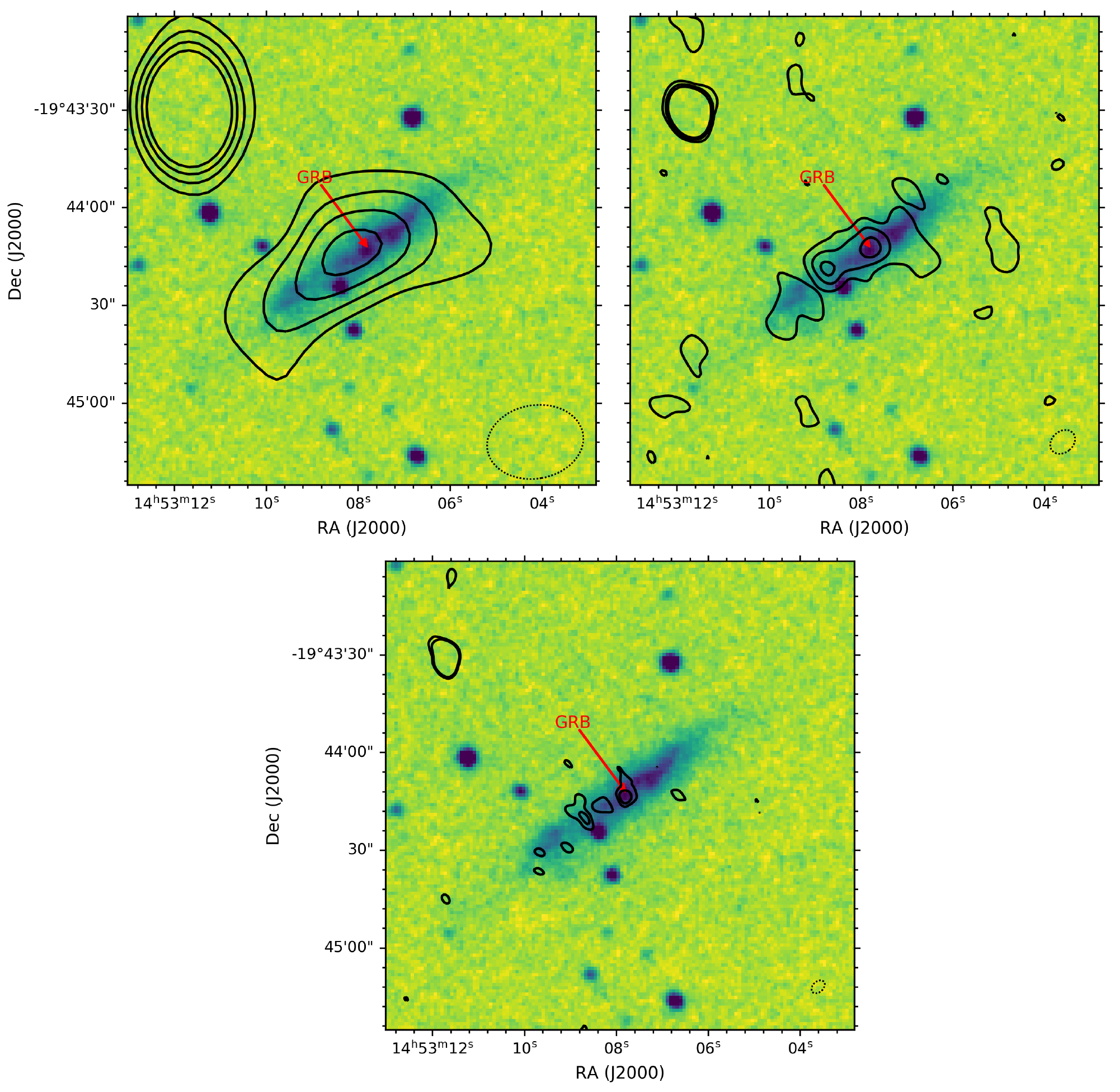}
\caption{\revone{1.4\,GHz continuum emission of the {\grb} host (black contours) at three different resolutions: 29.7\arcsec$\times$22.5\arcsec (top left), 8.4\arcsec$\times$6.5\arcsec (top right), and 4.7\arcsec$\times$3.2\arcsec (bottom; the beams are shown as grey dotted ellipses).
The contours are \revone{$3$, $5$, $7$, and $9\sigma$}
, where \revone{$\sigma$ = $176$ 
, $66$ 
and $40$ 
$\,\mu\mbox{Jy\,beam}^{-1}$} for low-, medium-, and high-resolution images, respectively.
The background is the UV image from \citet{Michalowski2018grb}. The position of {\grb} is marked by the arrow. The image size is 2.5\arcmin$\times$2.5\arcmin corresponding to 40.5 kpc $\times$ 40.5 kpc.}}
\label{continuum}
\end{figure*}

\begin{figure*}
\includegraphics[width=\textwidth]{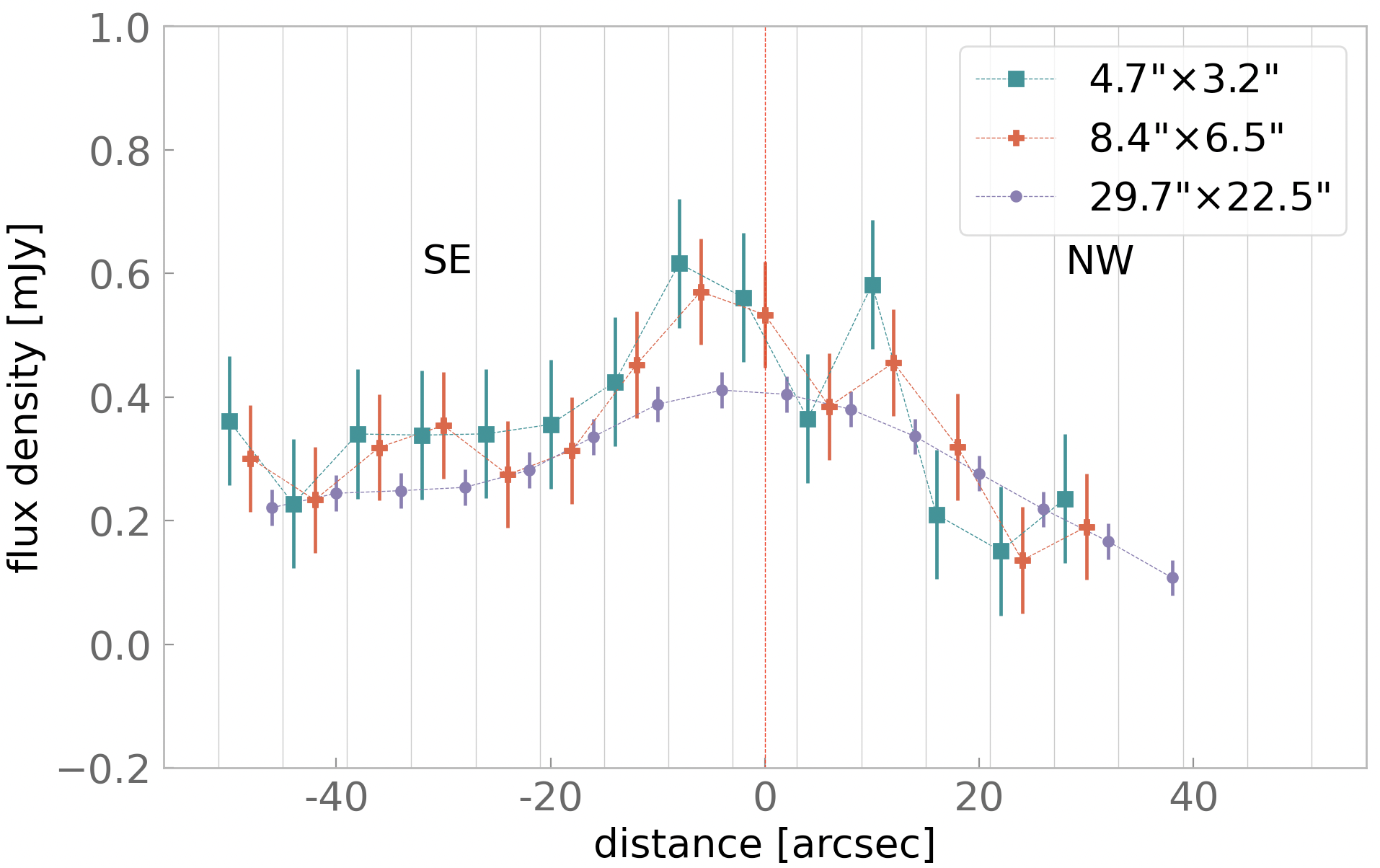}
\caption{\revone{1.4\,GHz continuum emission profiles of the {\grb} host, using the maps with a resolution of 29.7\arcsec$\times$22.5\arcsec, 8.4\arcsec$\times$6.5\arcsec, and 4.7\arcsec$\times$3.2\arcsec.
}}
\label{conProfiles}
\end{figure*}

\begin{figure*}
\includegraphics[width=\textwidth]{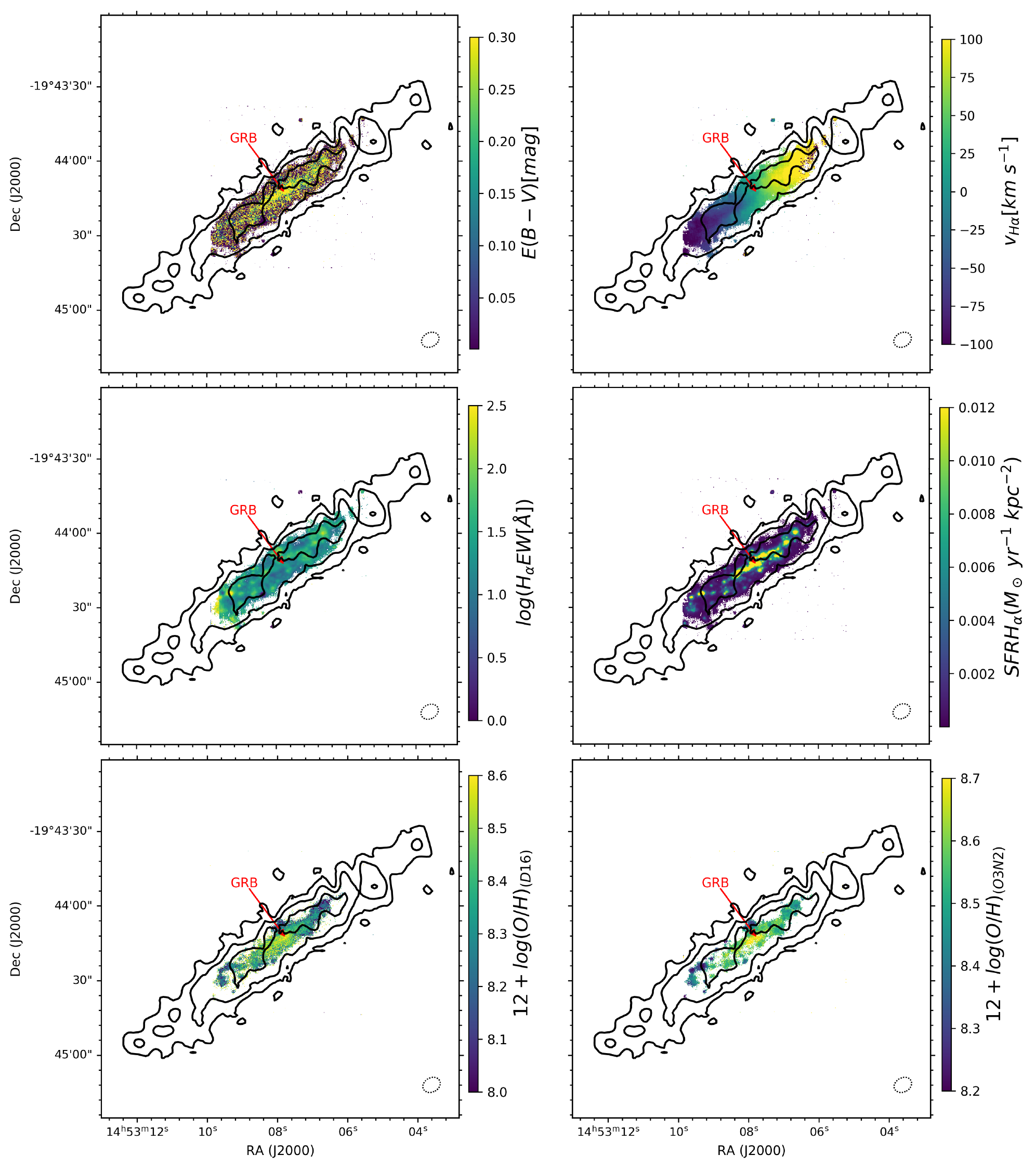}
\caption{\revone{Atomic gas emission of the {\grb} host from GMRT with a resolution of 7.4\arcsec$\times$5.7\arcsec (black contours) overlayed on maps of physical properties derived from MUSE observations by \citet{Tanga2018}.
(Top left) Dust reddening \textit{E(B-V)}.
(Top right) Velocity map (1st moment) based on the H$\alpha$ line, with respect to the redshift \textit{z} = 0.01326.
(Middle left) Equivalent width of H$\alpha$.
(Middle right) Star formation rate surface density derived from the H$\alpha$ emission.
(Bottom left) Metallicity using the diagnostic of \cite{Dopita2016}.
(Bottom right)  Metallicity based on the O3N2 method \citep{Pettini2004}.
The position of {\grb} is marked by the red arrow.
The beam of {\hi} observations is shown as a grey dotted ellipse.
The image size is 2.5\arcmin$\times$2.5\arcmin corresponding to 40.5 kpc $\times$ 40.5 kpc.}}
\label{muse}
\end{figure*}

\section{Tilted ring modelling}
\label{sec:model}

Due to the projection effects that are caused by the almost edge-on ($i$ $>$ 85$^{\circ}$) orientation of ESO 580-49 the rotation curve cannot be extracted from the velocity field. Therefore, in order to get  a better understanding of the distribution and dynamics of the {\hi} in ESO 580-49 we fit a tilted ring model \citep{Rogstad1974} to the GMRT observations. The Fully Automated TiRiFiC \citep[FAT, ][]{Kamphuis2015} fits the titlted ring model directly to the data cube and thus can also accurately model the observed {\hi} of discs with high inclination \citep{Jozsa2007}.

FAT fits all the parameters relating to the orientation and position of the disc. As such it determines the central coordinates and scale height of the disc as a whole for the model and allows for radial variations in the position angle (PA), inclination, rotational velocity, dispersion and surface brightness profile. We run FAT on the different resolution data cubes independently. 

For the highest resolution data (FWHM = 4.0\arcsec $\times$ 2.8\arcsec) FAT was unable to find enough flux in the data cube to reliably initiate the fitting process and hence no model was created. The lower resolutions were successfully fitted and all parameters are consistent between the two models \revone{at medium and low resolutions}. No radial variations are found in the dispersion, inclination, and PA.

\section{Results}

Our GMRT {\hi} data, the measurements made by MUSE \citep{Tanga2018} and the ultraviolet (UV) image \citep{Michalowski2018grb} of the host galaxy of {\grb} are presented in Figures \ref{HI}--\ref{MuseProfiles}. We show the intensity maps as well as profiles  along the galaxy. The profiles were constructed for all maps by measuring the signal in rectangular apertures perpendicular to the galaxy disk with a width of 6{\arcsec} and height of 17{\arcsec}, encompassing together the entire detectable emission.
\revone{Fig.~\ref{greyscale} in the appendix show the color-scale images of the moment 0 and continuum maps.}

The host has a clear S-shaped (`integral sign') warp in the UV image, which was not discussed in \citet{Michalowski2018grb}.

\subsection{\revthree{Residual scaling correction}}

\revtwo{The {\hi} and radio continuum fluxes in a given region appears to increase with increasing resolution of the cube. This is counter-intuitive, because usually better resolution leads to more flux being missed. First, we stress that for the most of the apertures the increase is less significant than $1\sigma$ and the maximum significance is only $2\sigma$. We however investigated the reason of this behaviour. 
In order to confirm that the masking does not introduce biases we performed cleaning at all resolution with a fixed masked defined by the emission from a low-resolution cube tapered at 5 k$\lambda$ and by just drawing a rectangle around the visible emission. In both cases we obtained similar results to our original analysis.
We also repeated the analysis without the {\sc TCLEAN}'s multiscale clean algorithm, again obtaining similar results.}

\revtwo{We conclude that the flux increase with resolution is due to a higher fraction of uncleaned flux and stronger sidelobes in the dirty beam at higher resolutions.  This makes the conversions from Jy beam$^{-1}$ to Jy pixel$^{-1}$ increasingly uncertain, because the uncleaned emission is not distributed according to a clean Gaussian beam with which the conversion is defined. We verified that this is the case by noting that the flux increase towards the highest resolution is proportional to  the difference between the areas of the dirty and cleaned beams.
}

\revthree{Hence, all {\hi} and radio continuum flux measurements were corrected for this effect using the method presented in \citet{Novak2019,Novak2020}, which is based on scaling the residual map by the  ratio of areas of the clean and dirty beams. For a given map and aperture size, this ratio can be found as $\epsilon=C/(D-R)$, where $C$, $D$, and $R$ are the measurements in this aperture with the clean component map, the dirty image and the residual image. Then the true flux is $C+\epsilon R$. In this way the residuals are scaled so that their unit matches that of the clean component map.}

\revthree{As expected, in this way the difference between fluxes at different resolutions decreased to a level less than $1\sigma$ for all the measurements except of the difference between the {\hi} measurements at the coarsest and middle resolutions for the circular apertures (Table~\ref{tab1}). The difference is less than $0.5\sigma$ for the same cubes for larger apertures encompassing the entire galaxy (the last row of Table~\ref{tab1}) or halves of the galaxy (Table~\ref{tab2}), so it is likely because the small circular apertures (matching the CO apertures) are too small and comparable with the beam size at the coarsest resolution, so the residual correction method works less efficiently for them. Hence, the {\hi} flux measurements in small circular apertures for the low-resolution cube are less reliable.}

\subsection{Atomic and molecular gas}\label{Atomic and molecular gas}

The left panels of Fig.~\ref{HI} show the contours of the {\hi} line (moment 0 maps) at three different resolutions overlayed on the UV image.
Atomic gas is smoothly distributed in the disk without any significant off-centre concentrations, plumes, or extra-planar gas, which could have suggested a recent inflow/outflow of gas or environmental interactions.
Specifically, there is no concentration of atomic gas at the GRB position.
There is however \revthree{$\sim$15\% more} atomic gas in the SE part of the galaxy, which is visible at all resolutions. \revone{The ratios of the atomic gas mass in the SE and NW halves are \revthree{1.16 $\pm$  0.04 (14.85\arcsec), 1.15 $\pm$ 0.04 (6.55\arcsec), and 1.14 $\pm$ 0.05 (3.4\arcsec).}}
The difference in the extent of the UV emission and atomic gas is clearly visible, with the atomic disk extending 
beyond the UV disk, which is common for spiral galaxies \citep[e.g.][]{Wang2013}.
 
The right panels of Fig.~\ref{HI} show the intensity weighted mean velocity maps. The velocity pattern is regular, as expected for a non-disturbed rotating disk.

Fig. \ref{HIProfiles} presents the profiles of the {\hi} emission along the galaxy. The top panel shows the {\hi} mass distribution, which is relatively symmetric, but in the NW part the {\hi} profile falls off faster. The lower panel presents the rotation curves  derived from the mean velocity (moment 1) maps.
We explored this to search for any irregularities in the moment 1 maps, which we did not find. We note that the rotation curves derived from the tilted ring modelling (see below) are more accurate. The full velocity width is around 250\,\kms.

\cite{Michalowski2018co} analysed the CO distribution in three regions in the host galaxy (see their Fig. 1) and we performed analysis of  atomic gas and properties derived from MUSE data at the same locations.
The total signal from each region was measured using the Starlink package \citep{Currie2014} 
with an aperture radius of 13.6{\arcsec}, matching the beam size of the CO observations. From the total {\hi} fluxes we measured the {\hi} masses ($M_{HI}$) based on eq.~2 in \cite{Devereux1990}. The positions of the apertures on our {\hi} map is shown on Fig.~\ref{apertures} in the appendix.
We also used an elliptical aperture encompassing the entire galaxy \revone{to calculate the total atomic gas mass ($\mbox{M}_{\rm HI,tot}$)}.
Table \ref{tab1} lists the results of these aperture photometry measurements.

The distribution of atomic gas is slightly asymmetric: there is around 15\revthree{$\pm$5\%} 
more gas in the SE region than in the the NW region. This was suggested by \citet{Michalowski2018grb} based on the asymmetry of the {\hi} line shape. 
We also calculated the molecular gas fraction using the molecular masses obtained by \cite{Michalowski2018co}. The SE region has \revone{twice} lower molecular gas mass and higher atomic gas mass than the NW region, so the molecular gas fraction is low \revthree{[$M_{\rm H2}/M_{\rm gas}\sim   (8.5\pm2)\%$, or $M_{\rm H2}/M_{\rm HI}\sim0.1$, see Table~\ref{tab1}]}.
\revone{This region encompasses one third of the host, but such low ratios are only found at the very outskirts of other galaxies  \citep[][their Fig.~17]{Leroy2008}.}

For the entire galaxy the atomic hydrogen mass was calculated using an elliptical aperture encompassing all the detected emission. The value measured here is \revone{consistent (M$_{\rm HI,tot}/\mbox{M}_{\rm HI,tot,single-dish}$ \revthree{$\sim(97\pm4)\%$)} with}
the measurement reported in \citet{Michalowski2015} based on the single-dish spectrum from \cite{Theureau1998}, \revone{especially taking into account the 10\% flux calibration uncertainty}. Fig. \ref{spectrum} shows the comparison of this single-dish HI spectrum and our GMRT observations at three spatial resolutions.
\revone{The spectra are consistent with each other indicating that GMRT did not miss any extended emission.}

\subsection{Kinematic modelling}
\label{sec:res:kin}

Table~\ref{tab3} presents the  properties of the {\grb} host based on the tilted ring modelling (Sect.~\ref{sec:model}) of low- and medium-resolution {\hi} cubes. Inclination values confirm that the model is consistent with a nearly edge-on orientation. The position of the galaxy centre determined in  the modelling is consistent with the position of {\grb} within 1--2$\sigma$. The GRB position is 0.1{\arcsec}  and 2.1{\arcsec} away from the centre in the low-resolution cube in right ascension and declination, respectively, whereas the errors of the centre position are 0.2{\arcsec} and 2.5{\arcsec}. For the medium-resolution cube the distance is 0.2{\arcsec}  and 2.8{\arcsec} and the errors are 0.1{\arcsec} and 1.1{\arcsec}.

The rotation curves derived from the tilted ring modelling are shown in Fig. \ref{3Dmodel} for data with resolutions of
16.8\arcsec$\times$12.9{\arcsec} and 7.4\arcsec$\times$5.7{\arcsec}. 
They present a flattening at large radii, typical for spiral galaxies \citep[e.g.][]{Rubin1980}, which signals a dominant dark matter contribution.
Based on the maximum rotational velocity we calculated a total dynamical mass of the galaxy of 10.8$\times$10$^{10}$ $M_\odot$ (the 14.85{\arcsec} data) and 9.7 $\times$10$^{10}$ $M_\odot$ (the 6.55{\arcsec} data). Subtracting  the total {\hi} mass from Table~\ref{tab1}, H$_2$ mass from \cite{Michalowski2018co}, and stellar mass from \cite{Michalowski2018grb}, we estimated a dark matter mass of 9.7$\times$10$^{10}$ $M_\odot$ (14.85\arcsec) and \revone{8.6$\times$10$^{10}$ $M_\odot$ (6.55\arcsec)}. This corresponds to a dark matter fraction around 90$\%$, so this galaxy is dark-matter dominated.

The ratios of the {\hi} masses included in these models and the {\hi} masses measured directly from the GMRT data show that the model is able to explain \revthree{$(83\pm3)\%$ and $(70\pm3)\%$}  of the emission in low- and mid-resolution cubes, respectively. We also compared the {\grb} host with the very tight relation between M$_{HI}$ and D$_{HI}$ of \citet[][their eq. 2]{Wang2016}. With a diameter of 178.9\arcsec and 158.8\arcsec \revone{(after deconvolving the beam, at a surface density of 1 $M_\odot$ pc$^{-2}$)}, this relation predicts the {\hi} mass of 6.9$\times$10$^{9}\,M_{\odot}$ and 5.4$\times$10$^{9}\,M_{\odot}$, respectively, \revone{factor of $1.33\pm0.04$ and $1.03\pm0.03$ above the measured value}. Hence, the host galaxy is 0.13 or 0.02\,dex below this relation. The scatter of the relation is 0.06 dex, so the {\grb} host galaxy is located at most 2 sigma away from it.

\subsection{Radio continuum}

Radio continuum emission at 1.4\,GHz presented in  Fig.~\ref{continuum} reflects the star formation rate in the host galaxy and has a similar extent to the UV emission. To first order, the maps at the three different resolutions show a symmetrical structure. However, they all show slightly more emission to the SE of the galaxy,
with \revthree{$(64\pm54)\%$}
more in the SE part at the lowest resolution.

Radio 1.4\,GHz continuum profiles are shown in Fig. \ref{conProfiles}. The data show a \revone{hint of} asymmetry at all resolutions, with the peak brightness shifted to the SE from the optical galaxy centre and the emission extending further in the SE direction. 
\revone{However, the sizes of error bars and the beam sizes prevent drawing a definite conclusion on this.}

\subsection{Stars, ionised gas and dust}

\revone{The H$\alpha$ SFR is corrected for dust extinction assuming the \citet{Calzetti1994} attenuation law with Rv=4.05. It is based on the H$\alpha$ flux to SFR conversion described in Kennicutt (1998) but assuming a Chabrier IMF (Chabrier 2003), which reduces the SFR by a factor of $\sim$1.7 compared to a Salpeter IMF.}

Regardless of the differences in gas content, the H$\alpha$-based SFRs in the NW and SE regions are very similar (Table \ref{tab1}). 
\revone{Using a  hybrid calibration combining H$\alpha$ with IR luminosities \citep{Kennicutt2012}, we obtained an SFR of $0.43\,M_\odot\,\mbox{yr}^{-1}$, which is consistent with that presented by \cite{Michalowski2018grb} based on infrared luminosity.}
The total SFR determined from the H$\alpha$ alone is \revone{$0.19\,M_\odot\,\mbox{yr}^{-1}$}, two times smaller than the IR-based SFR determined by \cite{Michalowski2018grb}. 
Similarly, the total radio continuum flux (Table~\ref{tab2}) implies an SFR of $0.15\,M_\odot\,\mbox{yr}^{-1}$ \citep[with the callibration of][]{Bell2003}.
The factor of two difference between these estimates and the IR SFR is within the typical calibration uncertainty, but indicates a contribution of evolved stars to dust heating, making the IR estimate too large.
However, we cannot rule out the possibility that the H$\alpha$ estimate is underestimated due to dust extinction and the radio estimate is underestimated due to age effects.

Fig.~\ref{muse} shows \revone{physical quantities derived from MUSE data \citep{Tanga2018} superimposed on the {\hi} moment zero map.} 
\revone{We show the full $1'$ extent of the data, whereas \citet{Tanga2018} showed the central $20"$. We present the SFR and the O3N2 metallicity maps for the first time. We infer the following results from these data.}
Similarly to the {\hi} observations, the H$\alpha$ velocity map shows a regular rotational pattern, \revone{also noted by \citet{Tanga2018}}.
Dust reddening \textit{E(B-V)} is highest along the middle of the disk, as expected given the prominent dust lanes \citep[Fig.~15 of][]{Michalowski2018grb}.
These central regions are also much more active than regions further from the disk plane, given the SFR surface density map
 based on the H$\alpha$ emission.
The H$\alpha$ equivalent width is a proxy of the age of a stellar population in young ($\sim10$\,Myr) starbursts (\citealt{Stasinska1996}; \citealt{Fernandes2003}). There are several regions with higher equivalent widths (\revone{$\sim200\AA$} and resulting lower ages), mostly at both ends of the disk and \revone{$3.5\arcsec$ (1\,kpc)} north-east of the centre \revone{(the latter was discussed by \citealt{Tanga2018})}.
Two methods were used by \citet{Tanga2018} to measure metallicity in the host galaxy.
Both diagnostics show a typical  radial behaviour with metallicity declining towards the outskirts. The region north-east of the galaxy centre does not fit into this pattern because it has as low metallicity as the regions at a similar height above the disk, but located much further from the centre ($12 + \log(O/H)$ = 8.1--8.2).
The diagnostic of \cite{Dopita2016} resulted in the NW region being the most metal-poor, but this is not reflected in the O3N2 method \citep{Pettini2004}.
Consistently with each other, both methods result in low metallicity for the SE part.

Fig. \ref{MuseProfiles} shows the profiles of six parameters calculated based on the MUSE data. \revone{Such analysis was not done by \citealt{Tanga2018}}. Dust reddening, $E(B-V)$, shows the largest value in the centre of the host galaxy decreasing with distance. The NW region is more reddened, with $E(B-V)$ decreasing less than in the SE half. The H$\alpha$ velocity profile is very symmetrical. The H$\alpha$ equivalent width is relatively flat, but the outermost region towards the SE reaches 150\,{\AA} and clearly stands out from the other locations. The highest value of the star formation density is found in the centre of the galaxy, near the {\grb} position, generally declining outwards. For metallicities derived with  both the D16 and O3N2 methods, the highest value occurs in the second aperture, $6\arcsec$ from the centre towards SE. This behaviour seems similar to the case of the 1.4\,GHz profiles presented in Fig.\ref{conProfiles}. This is because of the low-metallicity region north-east of the galaxy centre, lowering the average metallicity in the central aperture. An asymmetry is also evident in both metallicity profiles. The D16 metallicity in the SE half is higher than in the NW half, while we see the opposite for the O3N2 method.

We made similar measurements as in Table \ref{tab1} to investigate the differences between the two halves (SE and NW) of the galaxy. In Table \ref{tab2} we show the results of the aperture photometry on GMRT and MUSE maps with two apertures together encompassing the entire galaxy. There is more atomic gas in the SE half and this is true for each resolution, \revone{but the difference is significant only at a $\sim2\sigma$ level. A similar difference is apparent for the radio continuum flux.} 
The average dust reddening for both halves is similar, unlike what the profile suggests (Fig.~\ref{MuseProfiles}). This is because the average value is dominated by a large number of pixels with low $E(B-V)$. Due to regions with high H$\alpha$ equivalent widths in the SE end of the galaxy, the corresponding half has a high average value of this parameter. The total H$\alpha$ SFRs of the galaxy halves are comparable. In the case of metallicity, we again see differences in results depending on the method used. For D16 the average metallicity of the SE half is higher, whereas the O3N2 method results in similar values for both halves.

\subsection{Large-scale environment}

We also analysed the large-scale environment around the {\grb} host using the NASA/IPAC Extragalactic Database (NED). The closest galaxies are WISEA J145239.17-192125.8 and ESO 580-G052, 24{\arcmin} and 30{\arcmin} (321 kpc and 301 kpc) away, respectively. Within 1\,Mpc in projection and $\Delta v=~\pm1500~\kms$ there are 6 galaxies. Moreover, 1.37 Mpc (1.4 deg) away from the {\grb} host there is a galaxy group with NGC 5791 as the brightest member \citep{Crook2007, Diaz-Gimenez2012}. The {\grb} host appears to lie in the outskirts of this group. 

We note that the SN-less GRB\,060505 was found 4\,Mpc from a galaxy cluster \citep{Thone2008}, whereas the relativistic SN\,2009bb was found 600\,kpc from a galaxy group \citep{Michalowski2018sn}. The small number of such studies prevents drawing firm conclusions.

\section{Discussion}

\begin{figure*}
\includegraphics[width=0.85\textwidth]{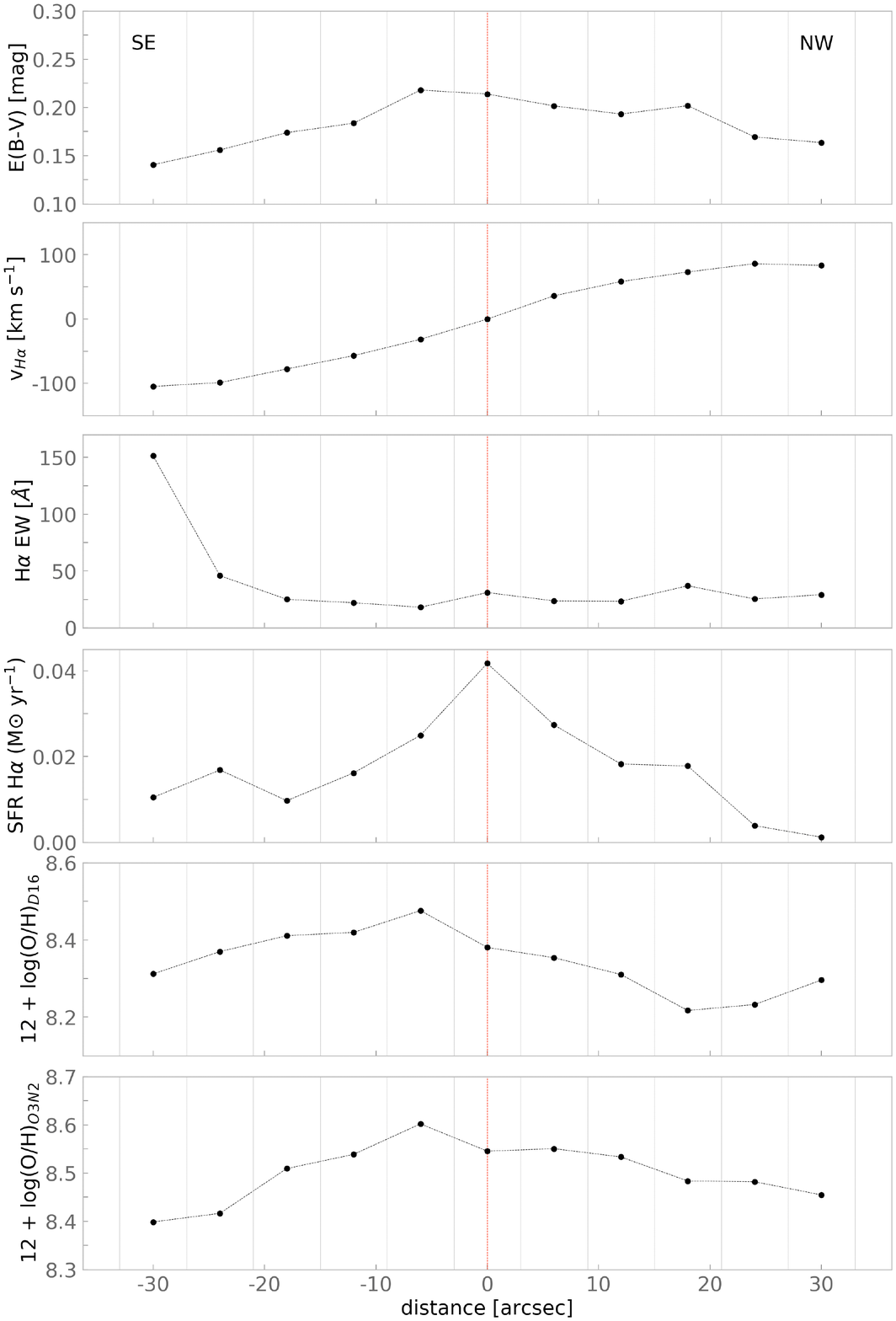}
\caption{Profiles of properties derived using the MUSE data (Fig. \ref{muse}) as a function of the distance from the galaxy centre (red dashed line).}
\label{MuseProfiles}
\end{figure*}

The host of {\grb} displays many regular features: largely symmetrical atomic gas (Fig.~\ref{HI} and \ref{HIProfiles}, Table~\ref{tab2}) and radio continuum (Fig.~\ref{continuum} and \ref{conProfiles}, Table~\ref{tab2}) distributions, and rotational patterns derived from both {\hi} and H$\alpha$ lines (Fig.~\ref{muse} and \ref{MuseProfiles}). There are only small deviations from this regularity, which will be discussed below.

Hence, there is no evidence of strong and recent gas inflow/outflow or environmental interactions.
This is different from the irregular ISM distributions in the hosts of GRBs, 
\revone{broad-lined type Ic (Ic-BL) SNe, \revthree{and fast radio bursts (FRBs)}. Four such explosions were found near the most significant concentration of atomic gas: GRB 980425 \citep{Arabsalmani2015,Arabsalmani2019,Michalowski2015}, GRB 060505 \citep{Michalowski2015},  SN\,2009bb \citep{Michalowski2018sn}, and SN\,2002ap \citep{Michalowski2020}.
Moreover, GRB\,980425 \citep{Michalowski2014,Michalowski2016}, GRB\,060505 \citep{Thone2008,Thone2014}, GRB\,100316D \citep{Izzo2017}, and SN\,2009bb \citep{Michalowski2018sn} exploded close to the region that is the brightest in the infrared, radio, H$\alpha$, and [OI].
\revthree{Similarly, the {\hi} line profiles of two FRB hosts for which such measurement is possible were found to be extremely asymmetric, compared to those of the general population of galaxies \citep{Michalowski2021}.}
On the other hand the ISM was found to be more regular for the host of SNe type Ib \citep{Michalowski2020b} and the transient AT\,2018cow \citep{Michalowski2019}.
\revfour{However, \cite{Roychowdhury2019} reported AT\,2018cow in a distorted dense {\hi} ring-like structure.}
}

\revone{Most long nearby GRBs for which deep spectral observations are possible are associated with SNe type Ic-BL \citep{Hjorth2012}, exhibit sub-solar metallicities in their environments \citep{Leloudas2011,Modjaz2011,Japelj2016}, and exploded in galaxies with irregular ISM distributions (see above). On the other hand, {\grb} was not associated with an SN, exploded in a region with solar metallicity, and its host has a regular ISM distribution. These properties may suggest that the progenitor of {\grb} is different from that of most long GRBs.}
This could be a compact object merger \citep{Wang2017,Michalowski2018grb,Tanga2018,Yue2018,Dado2018}.
Indeed, the host galaxy of the neutron star merger GW\,170817 has  a regular ISM distribution with only minor irregularities  \citep{Levan2017}. 
\revone{Low numbers of GRB hosts with characterised ISM properties precludes drawing conclusions from the ISM distribution alone.
}

There are a few more subtle irregularities of the {\grb} host, which may point to a weak gas inflow or interaction.
\revone{We note that these irregularities are weak and the significance of most of them is around $2\sigma$, so deeper observations are needed to investigate this topic. However, these irregularities were found in independent datasets, so their combined significance is higher.}
\revthree{We also note that these features can in principle be explained by different mechanisms.}

The S-shape of the galaxy, clear on the UV image (Fig.~\ref{continuum}), \revone{$2\sigma$} asymmetry in the 1.4\,GHz continuum image (Figs.~\ref{continuum} and \ref{conProfiles}, Table~\ref{tab2}), and the asymmetric {\hi} profile (Fig.~\ref{spectrum}) suggests an interaction.
S-shaped warps in stellar  \citep{Reshetnikov1999,Reshetnikov2016,Ann2006} and {\hi} \citep{Sancisi2008} distributions are common and were claimed to be indications of tidal interaction or gas accretion. Similarly, S-shapes in simulations have been induced by interactions, or by gas accretion, and can be sustained several Gyr after such events (\citealt{Kim2014,Gomez2016,Gomez2017,Semczuk2020}). \revone{Such a long timescale makes the GRB event unlikely to be connected with a potential interaction.} Ram pressure stripping can also induce S-shaped warps \citep{Haan2014}, but this probably does not apply to the {\grb} host because it is more than 1\,Mpc away from a galaxy group, so the intragroup medium density is unlikely to be high at its position and any infall velocity must be low.

Asymmetric {\hi} spectrum (\citealt{Michalowski2018grb}; Fig.~\ref{spectrum}) resulting from more atomic gas in the SE part of the galaxy (Table~\ref{tab2}) also indicates external influence, e.g. tidal interaction or gas inflow.
Indeed, \citet{Watts2020} found that galaxies with  asymmetric {\hi} line profiles generally contain 29\% less {\hi} than their symmetric counterparts, and this is due to gas removal during the interaction with galaxy environment \citep[see also][]{Reynolds2020,Hu2021}. 
\revone{Indeed, the host of {\grb} contains ($35\pm4$)\% less atomic gas than predicted from the relation between the atomic gas and size (see Sect.~\ref{sec:res:kin}).}

\revone{We quantified the {\hi} line asymmetry of the {\grb} host using the diagnostics defined by \citet{Reynolds2020}. The {\grb} host has a difference between the flux-weighted velocity and the midpoint velocity at 50\% of the peak flux of $\Delta V_{\rm sys}=2.6\,\kms$ (eq.~4 in \citealt{Reynolds2020}); the integrated flux ratio between the left and right halves of the spectrum of $A_{\rm flux}=1.24$ (eq.~6); the flux ratio between  the left and right peaks of the spectrum of $A_{\rm peak}=1.31$ (eq.~7); and the residual from the subtraction  of the spectrum flipped around the flux-weighted velocity from the original spectrum of $A_{\rm spec}=0.26$ (eq.~8). We compared these values with measurements for galaxies with stellar masses of $9<\log(M_{\rm star}/M_\odot)<10$ from the Local Volume {\hi} Survey (LVHIS; \citealt{Koribalski2018}) and the VLA Imaging of Virgo in Atomic Gas (VIVA; \citealt{Chung2009}), with low and high environmental density respectively (see Fig.~8 and Table~3 of \citealt{Reynolds2020}). All asymmetry diagnostics for the {\grb} host, except $A_{\rm spec}$, are 1--2 standard deviations higher than the mean for the low-density LVHIS galaxies and consistent with the mean for the high-density VIVA galaxies. Hence, the {\hi} spectrum of the {\grb} host is at the highest end of asymmetry for galaxies in the low-density environments and is similarly asymmetric as spectra of galaxies in a cluster. This supports the hypothesis that this asymmetry is related to the interaction with the environment (gas inflow, merger, etc.).
}

The SE part of the galaxy also shows some features, which may suggest gas accretion. There is \revone{twice less} molecular gas \citep{Michalowski2018co} and more atomic gas than in the NW part, \revone{so the molecular gas fraction is approximately a factor of two lower, but this is significant only at a $2\sigma$ level} (Tables \ref{tab1} and \ref{tab2}). The radio continuum emission is also stronger in the SE than in the NW (Figs.~\ref{continuum}, \ref{conProfiles} and Table~\ref{tab2}), \revone{again significant at a $2\sigma$ level}. 
\revone{If these differences are confirmed with deeper data, then they can be explained by either interaction or gas inflow. Both processes can lead to higher atomic gas density in a part of a galaxy (either directly during inflow or by rearranging gas distribution during interaction). This can can also enhance the SFR at that position due to a higher gas density, explaining stronger radio emission.}
If this process was recent, then the {\hi} to H$_2$ conversion has not taken place yet. The metallicity effect could in principle explain low CO emission \citep{Bolatto2013}, but we measured a similar metallicity in the CO-rich north-western part (Fig.~\ref{muse}, \ref{MuseProfiles} and Table \ref{tab1}), so this is not the case for the SE region.

Similarly, the region just outside of the galaxy centre, \revone{3.5{\arcsec} (1\,kpc)} to the north-east has unusual properties \revone{(see Figs. 5--7 of \citealt{Tanga2018} for a zoom-in view of the MUSE data).} It 
has untypically low metallicity \revone{(8.15 for the D16 calibration and 8.25 for the O3N2 calibration)}, 
\revthree{compared to other regions at similar galactocentric distances,}
and exhibits high H$\alpha$ equivalent width \revone{($\sim200\AA$)}, suggesting a very young stellar population (Fig.~\ref{muse}).
These properties are consistent with gas flowing from the intergalactic medium and enhancing star formation \revthree{\citep[see also][]{sanchezalmeida13,sanchezalmeida14,sanchezalmeida14b,sanchezalmeida15}}.
\revone{However, only the detection of a stream of gas extending outside the galaxy would provide strong evidence supporting this conclusion.}
If this is correct, then the birth of the progenitor of {\grb} may be related to this process. 

On the other hand, we did not find any definitive  signature of outflows.
The MUSE spectrum at the {\grb} position does not show any deviation from a single Gaussian profile.
Conversely, \citet{Thone2019} found that all long GRB hosts experience strong outflows, based on the existence of broad mostly blueshifted H$\alpha$ components, kinematically decoupled from narrow components. However, in our case, even if such an outflow is present, we would not be able to detect it in the spectra, because it would likely be directed perpendicular to the line-of-sight for this edge-on galaxy.

With our radio continuum image (Fig.~\ref{continuum}) we can also test for the presence of a radio-loud AGN. We did not detect a strong point source in the centre, so we rule out this possibility.

The reason for the discrepancy between the metallicity measurements using the D16 and O3O2 methods (Fig.~\ref{MuseProfiles} and Table \ref{tab1}) is likely the same as described in \citet{Kruhler2012}. Namely the O3N2 method misinterprets highly ionised regions for having low metallicity. The D16 method takes ionisation into account and therefore is more accurate.

Interaction with two nearby galaxies 300\,kpc away \revone{(in projection)} may be responsible for the S-shape of the {\grb} host, \revone{if in the past they were closer. The distortion may also be due to a past minor merger.} On the other hand the projected distance to the NGC 5791 galaxy group of 1.37 Mpc makes it unlikely that interaction with the group can significantly modify the distribution of gas or stars the {\grb} host. However, the proximity to the group implies the presence of a supply of intergalactic gas to be accreted by the {\grb} host. A similar situation was found for the SN\,2009bb host \citep{Michalowski2018sn}.
\revone{The projected virial radius of the NGC 5791 group is 0.24\,Mpc \citep[][based on the separations of the group members]{Crook2007}. Hot, X-ray emitting gas has been detected out to more than 0.5\,Mpc from group centres \citep{Rasmussen2007,Mernier2017}. Hence, \revthree{the {\grb} host is not a member of this group, but at its position, i.e.~at the distance of 3--4 times the group radius}, one can expect some gas from the group.
}

\begin{table*}
\caption{\revone{Properties of two parts in the {\grb} host (NW half and SE half).}}
\centering
\begin{tabular}{ccccccccccccc}
\hline\hline
\multirow{2}{*}{Reg.} & & \multicolumn{3}{c}{M$_{\rm HI}$} & \multicolumn{3}{c}{1.4\,GHz} & E$(B-V)$ & H$\alpha$EW & \multicolumn{2}{c}{$Z$} &  SFR \\ 
 & &  \multicolumn{3}{c}{(10$^9$ $M_\odot$)} & \multicolumn{3}{c}{(mJy)} & (mag) & ($\AA$) & \multicolumn{2}{c}{$12 + \log(O/H)$} & ($M_\odot$ yr$^{-1}$) \\
\cline{3-8}\cline{11-13}
 & & 14.85\arcsec & 6.55\arcsec & 3.4\arcsec & 26.5\arcsec & 7.45\arcsec & 3.95\arcsec &  &  & D16 & O3N2 &\\
\hline

NW half & & \revthree{2.58} & \revthree{2.66} & \revthree{2.83} & \revthree{1.37} & \revthree{1.78} & \revthree{1.90} & 0.19 & 29 & 8.31 & 8.52 & 0.09\\
 & \revone{$\pm$} & \revthree{0.09} & \revthree{0.09} & \revthree{0.12} & \revthree{0.36} & \revthree{0.45} & \revthree{0.33}\\
 
SE half & & \revthree{2.98} & \revthree{3.06} & \revthree{3.23} & \revthree{2.49} & \revthree{2.93} & \revthree{2.97} & 0.18 & 70 & 8.41 & 8.51 & 0.10\\
 & \revone{$\pm$} & \revthree{0.06} & \revthree{0.07} & \revthree{0.09} & \revthree{0.36} & \revthree{0.45} & \revthree{0.33} \\
\hline
\end{tabular}
\label{tab2}
\tablebib{Atomic gas masses (M$_{\rm HI}$), 1.4 GHz continuum emission, dust reddening E(B-V), equivalent width of H$\alpha$, metallicities determined from the \citet[][D16]{Dopita2016}  and \citet[][O3N2]{Pettini2004} methods, and H$\alpha$ based  star formation rates. The {\hi} properties and 1.4\,GHz continuum emission are given in three columns, corresponding to different resolutions, indicated in arcsec.}
\end{table*}

\begin{table*}
\caption{Properties of the host galaxy of {\grb} determined from {\hi} cubes at two resolutions, 14.85\arcsec and 6.55\arcsec, using tilted ring modelling.}
\centering
\begin{tabular}{cccccccccc}
\hline\hline
Res. & RA & Dec & incl. & v$_{\rm rot, max}$ & D$_{\rm HI}$ & D$_{\rm HI}$ & flux$_{\rm tot}$ & M$_{\rm HI, mod}$ & M$_{\rm HI, mod}$ / 
\\
 & (hh:mm:ss) & (dd:mm:ss) & ($^\circ$) & (km s$^{-1}$) & (arcsec) & (kpc) & (Jy km s$^{-1}$) & (10$^9$ $M_\odot$) &
 M$_{\rm HI}$
 \\
\hline
14.85\arcsec & 14:53:07.92 & -19:44:14.08 & 85 & 125 & 179 & 48.4 & 6.36 & 4.7 & \revthree{0.83$\pm$0.03}\\
6.55\arcsec & 14:53:07.98 & -19:44:14.75 & 83 & 129 & 159 & 43.0 & 5.48 & 4.0 & \revthree{0.70$\pm$0.03}\\

\hline
\end{tabular}
\label{tab3}
\tablebib{The coordinates of the galaxy centre (RA, Dec), inclination of the galaxy with respect to the line-of-sight (incl.), maximum rotational velocity (v$_{\rm rot, max}$), diameter of the {\hi} disc at a surface density of 1 $M_\odot$ pc$^{-2}$ (D$_{\rm HI}$), total {\hi} line flux density (flux$_{\rm tot}$), atomic gas mass in the model (M$_{\rm HI, mod}$), and ratio between M$_{\rm HI, mod}$ and the {\hi} mass derived directly from the GMRT data (M$_{\rm HI}$, see Table~\ref{tab1}).}
\end{table*}

\section{Conclusions}
Based on new GMRT and archival MUSE observations of the host galaxy of the unusual long {\grb}, we have characterized the interstellar medium properties of the host across the galaxy and in the vicinity of the GRB explosion. Deep observations of SN-less GRB hosts are scarce and hence this study contributes significantly to the understanding of ISM properties in such galaxies.

The host galaxy of {\grb} is characterized by regular largely symmetrical atomic gas, radio continuum distribution, and rotational patterns with only small deviations from this regularity. 
This is different from the irregular ISM distributions seen in the hosts of long GRBs and type Ic SN, which may suggest that the progenitor of {\grb} is different from the explosion of a very massive star, consistent with the fact that no SN was found to be associated with the GRB.

Subtle irregularities include  a warped S-shape in the 
UV image,
asymmetry in the {\hi} and radio continuum distribution, a low-metallicity region close to the GRB position, and a region with very high H$\alpha$ EW. This suggests weak interaction with inflowing gas or tidal forces with another galaxy.  
\revone{We note that these irregularities are weak and the significance of most of them is around $2\sigma$, so deeper observations are needed to investigate this topic. However, these irregularities were found in independent datasets, so their combined significance is higher.}
 
Two other galaxies are present within  300\,kpc and they can be responsible for the S-shape of the {\grb} host. Additionally there is a group of galaxies 1.37\,Mpc away, whose intergalactic medium may fuel frequent gas inflows into the GRB host.

\begin{acknowledgements}
We wish to thank the referee for a detailed comments, which helped us to clarify our conclusions. 
A.L. and M.J.M.~acknowledge the support of 
the National Science Centre, Poland through the SONATA BIS grant 2018/30/E/ST9/00208. 
This research was funded in whole or in part by National Science Centre, Poland (grant numbers: 2021/41/N/ST9/02662 and 2020/39/D/ST9/03078).
For the purpose of Open Access, the author has applied a CC-BY public copyright licence to any Author Accepted Manuscript (AAM) version arising from this submission.
M.J.M.~acknowledges the Fulbright Senior Award  from the Polish-U.S. Fulbright Commission.
P.K. is supported by the BMBF project 05A17PC2 for D-MeerKAT.
J.H. was supported by a VILLUM FONDEN Investigator grant (project number 16599).
L.K.H.~acknowledges funding from the INAF PRIN-SKA program 1.05.01.88.04. The Cosmic Dawn Center is funded by the DNRF.
M.P.K. acknowledges support from the First TEAM grant of the Foundation for Polish Science No. POIR.04.04.00-00-5D21/18-00.
This article has been supported by the Polish National Agency for Academic Exchange under Grant No. PPI/APM/2018/1/00036/U/001. 

We thank the staff of the GMRT who have made these observations possible. GMRT is run by the National Centre for Radio Astrophysics  of the Tata Institute of Fundamental Research. 
This research has made use of 
SAOImage DS9, developed by Smithsonian Astrophysical Observatory \citep{ds9}; 
the NASA/IPAC Extragalactic Database (NED), which is operated by the Jet Propulsion Laboratory, California Institute of Technology, under contract with the National Aeronautics and Space Administration;
and
the NASA's Astrophysics Data System Bibliographic Services.
We acknowledge the use of Starlink software which is currently supported by the East Asian Observatory.
We acknowledge the usage of the HyperLeda database (\url{leda.univ-lyon1.fr}). 

\end{acknowledgements}

\bibliographystyle{aa_like_apj}
\bibliography{main}

\appendix
\section{Color-scale maps}

Below we present color-scale images of the moment 0 and continuum maps.

\begin{figure*}
\includegraphics[width=\textwidth]{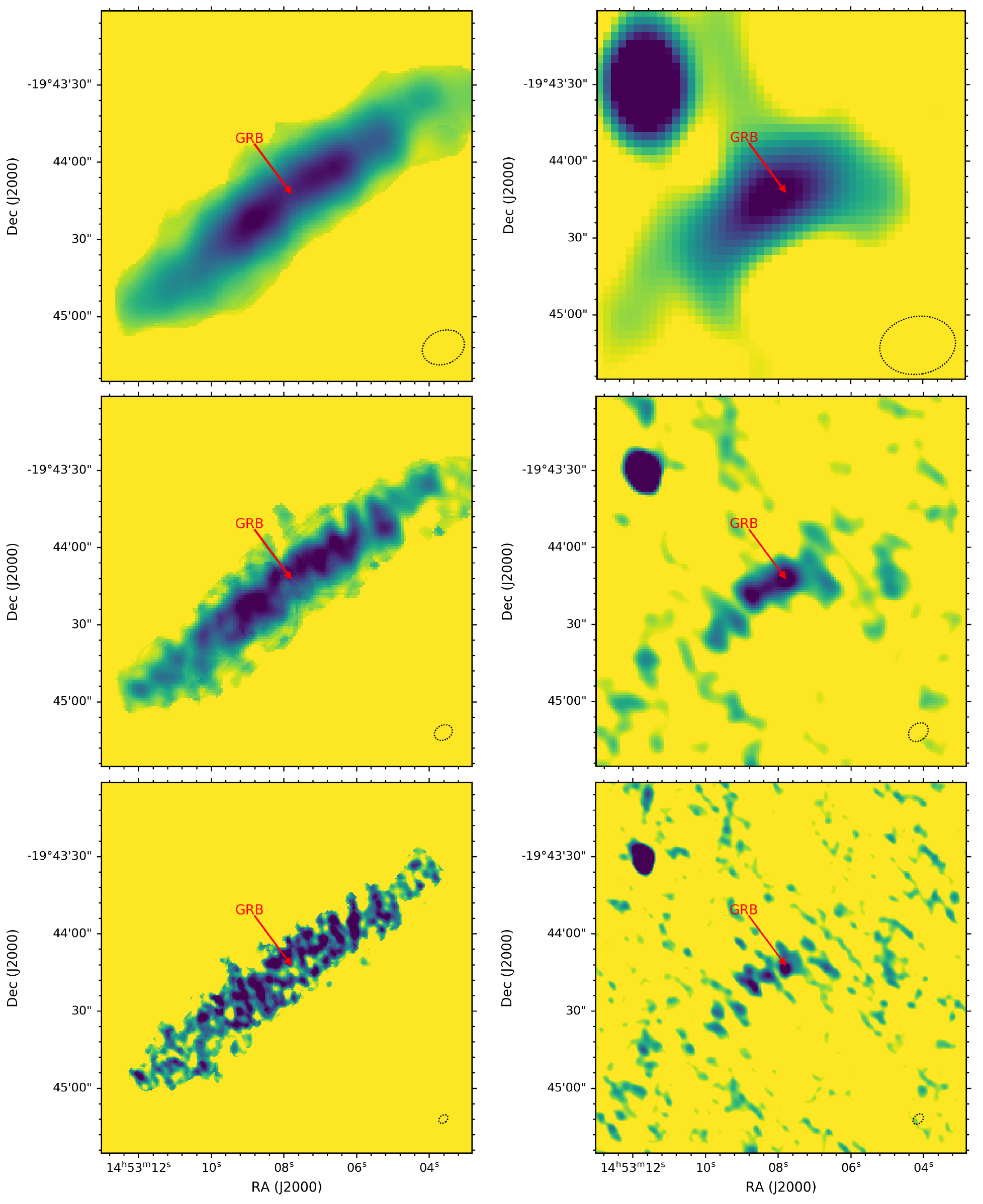}
\caption{\revone{(Left) Color-scale images of moment 0 maps of the {\grb} host detected by GMRT with three different resolutions from top to bottom: 16.8\arcsec$\times$12.9\arcsec, 7.4\arcsec$\times$5.7\arcsec, and 4.0\arcsec$\times$2.8{\arcsec} (the beams are shown as grey dotted ellipses). (Right) Color-scale images of 1.4\,GHz continuum emission of the {\grb} host at three different resolutions: 29.7\arcsec$\times$22.5\arcsec (top), 8.4\arcsec$\times$6.5\arcsec (middle), and 4.7\arcsec$\times$3.2\arcsec (bottom; the beams are shown as grey dotted ellipses).}} 
\label{greyscale}
\end{figure*}

\section{Positions of regions}

Below we show the positions of regions analysed in CO in \citet{Michalowski2018co}.

\begin{figure*}
\includegraphics[width=\textwidth]{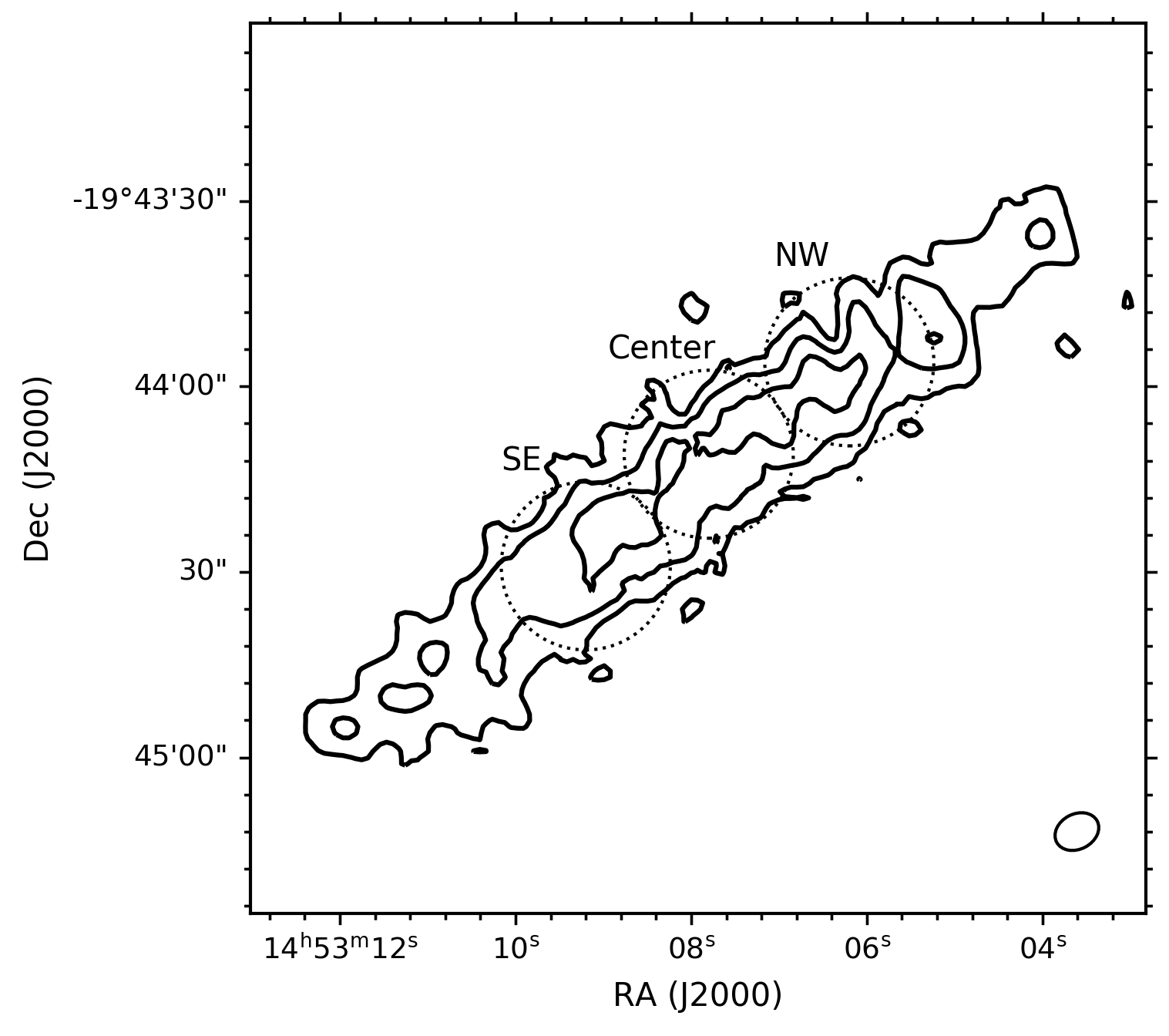}
\caption{\revone{Positions of regions analysed in CO in \citet[][dotted circles]{Michalowski2018co} on our {\hi} contours with a resolution of 7.4\arcsec$\times$5.7\arcsec. The radii are 13.6{\arcsec}, corresponding to the beam size of the CO observations. These apertures were used to calculate the properties shown in Table~\ref{tab1}.}}
\label{apertures}
\end{figure*}

\end{document}